\documentclass[twocolumn,superscriptaddress]{revtex4-2}

\usepackage[utf8]{inputenc}
\usepackage{amsmath}
\usepackage{graphicx}
\usepackage{float}
\usepackage{braket}
\usepackage{bbm} 
\usepackage{color}
\usepackage{siunitx}

\usepackage{subcaption}

\newcommand{\srr}{\hat{\sigma}^{rr}}
\newcommand{\sgr}{\hat{\sigma}^{gr}}
\newcommand{\srg}{\hat{\sigma}^{rg}}
\newcommand{\sgg}{\hat{\sigma}^{gg}}
\newcommand{\ddt}{\frac{\text{d}}{\text{d}t}}
\newcommand{\rfac}{r_\text{f}}
\newcommand{\gfac}{\Gamma_\text{f}}
\newcommand{\drfac}{\delta r_\text{f}}

\begin{document}

\begin{abstract}
    The excitation spread caused by Rydberg facilitation in a gas of laser driven atoms is an interesting model system for studying epidemic dynamics. We derive a mean-field approach to describe this facilitation process in the limits of high and low temperatures, which takes into account Rydberg blockade and the network character of excitation spreading in a low-temperature gas. As opposed to previous mean-field models, our approach accurately predicts all stages of the facilitation dynamics from the initial fast epidemic growth, an extended saturation period, to the final relaxation phase. 
\end{abstract}

\title{Mean-field approach to Rydberg facilitation in a gas of atoms at high and low temperatures.}

\author{Daniel Brady}
\affiliation{Department of Physics and Research Center OPTIMAS, University of Kaiserslautern-Landau, D-67663 Kaiserslautern, Germany}
\author{Michael Fleischhauer}
\affiliation{Department of Physics and Research Center OPTIMAS, University of Kaiserslautern-Landau, D-67663 Kaiserslautern, Germany}

\date{\today}

\maketitle

\section{Introduction}

Rydberg atoms have gained a lot of interest in the last few decades due to their strongly exaggerated properties. In particular, they have very long life times and strong interactions over  distances covering several µm \cite{gallagher2006rydberg}. These features allow Rydberg systems to be especially useful in a multitude of applications such as quantum information processing \cite{PhysRevLett.85.2208,PhysRevLett.87.037901,gaetan2009observation,urban2009observation,saffman2010quantum} or the study of many-body spin physics 
\cite{weimer2010rydberg,schauss2012observation,Bernien2017,Browaeys2020,Surace2020,scholl2021quantum,Leseleuc2019,Semeghini2021}.

One interesting process in many-body Rydberg systems is Rydberg facilitation, which has been used to study dissipative spin systems \cite{PhysRevA.98.022109}, transport and localization phenomena in disordered systems \cite{marcuzzi2017facilitation}, or self-organized criticality \cite{helmrich2020}.

In this type of many-body system, atoms are coupled off-resonantly to a Rydberg state. As a result of the Rydberg dipole interaction, however, atoms near an already excited Rydberg atom can be excited resonantly. Thus an initial seed excitation can lead to a cascade of excitations.
It has been shown experimentally  that this type of system bears close similarities to epidemic dynamics \cite{natcom_griffiths}. 

The most simple description of epidemic type systems is given by susceptible-infected-susceptible (SIS) models. Here,
each individual has two internal states, susceptible (S) or infected (I). Susceptible individuals are infected with rate $\kappa$ in the presence of other, already infected ones, and an infected individual can return to the susceptible state with rate $\gamma$ \cite{anderson1991infectious,bailey1975mathematical,murray1993epidemic}.

Under the assumption of homogeneous mixing, where all individuals interact with each other completely at random, all information about the epidemic dynamics is contained in the total
fractions $\rho^\nu$ in the susceptible ($\nu$=S) and infected state ($\nu$=I), which obey simple homogeneous mean-field 
equations given by \cite{pastor2015epidemic}
\begin{subequations}
    \label{eq:rhoNU}
    \begin{align}
        \label{eq:rhoI}
        \ddt \rho^\textrm{I} &= \kappa \rho^\textrm{I} \rho^\textrm{S} - \gamma \rho^\textrm{I},
        \\
        \label{eq:rhoS}
        \ddt \rho^\textrm{S} &= -\kappa \rho^\textrm{I} \rho^\textrm{S} + \gamma \rho^\textrm{I}.
    \end{align}
\end{subequations}
These systems feature an absorbing-state phase transition between two dynamical phases, namely an absorbing phase in which all infections die out, and an active phase where - in the thermodynamic limit - infections last forever. A suitable order parameter to distinguish these phases is the steady-state active (infected) density $\rho_\mathrm{ss}^\textrm{I}$. From eq.~\eqref{eq:rhoI} one recognizes that this phase transition occurs when
\begin{align}
    \label{eq:sis_phase_transition}
    \rho^\textrm{S}\kappa = \gamma.
\end{align}
%
\begin{figure}[H]
  \centering
  \includegraphics[width=\columnwidth]{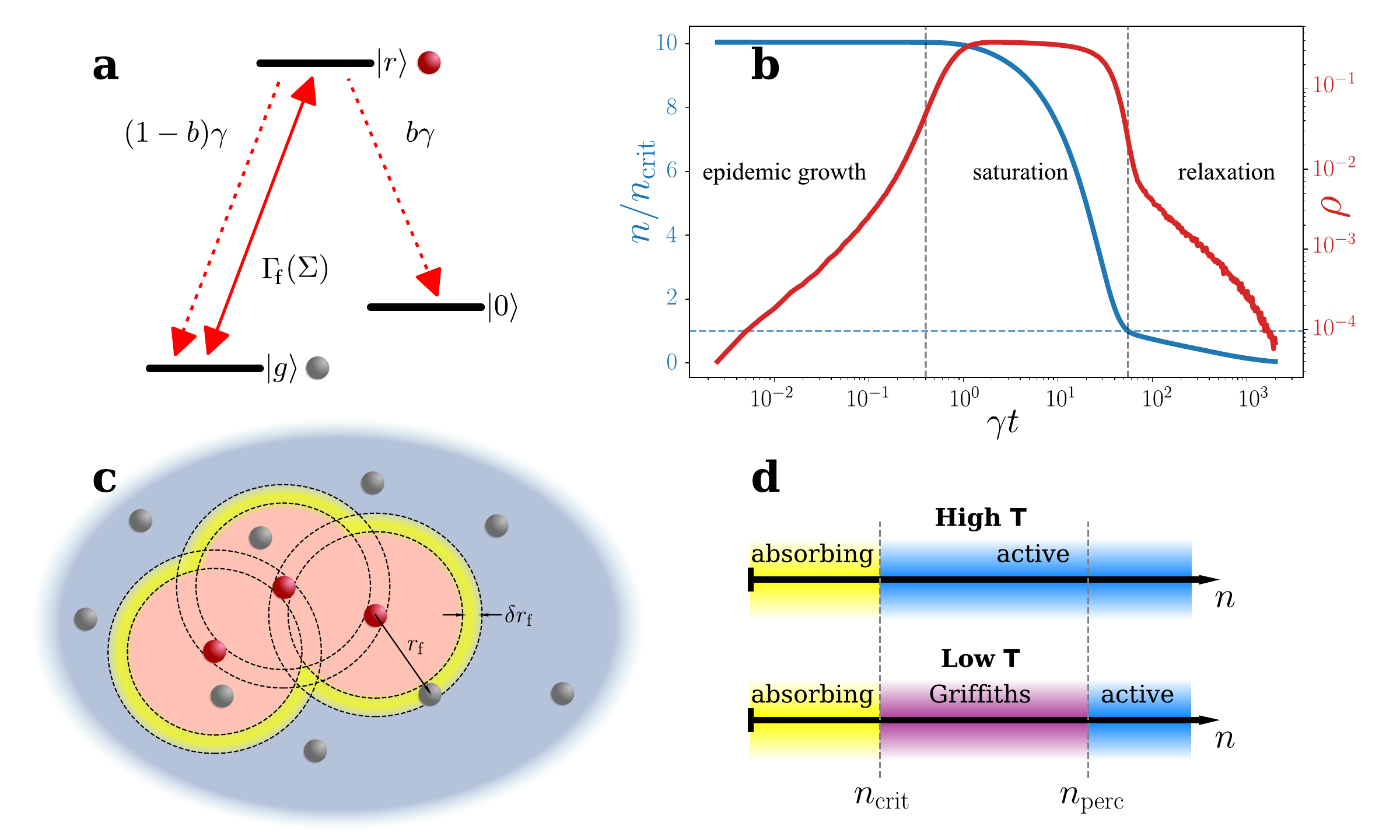}
  \caption{(a):~Level scheme of internal atomic states with ground $\ket{g}$ (susceptible) state, Rydberg $\ket{r}$ (infected) state, and inert $\ket{0}$ (recovered) state. An external laser drives the transition between $\ket{g}$ and $\ket{r}$ and spontaneous decay occurs from $\ket{r}$ to either $\ket{g}$ or $\ket{0}$, modulated by branching parameter ${b \in [0, 1]}$. (b):~Monte-Carlo data of the Rydberg density $\rho$ (red) and total active density $n$ (in states $\vert g\rangle$ and $\vert r\rangle$) (blue) over time, with ${b = 0.3}$ and initial condition ${\rho(t=0) = 0}$, showing the typical epidemic stages. The blue dashed line corresponds to ${n / n_\mathrm{crit} = 1}$. (c):~Schematic of Rydberg atoms (red dots) spanning facilitation shells (yellow region) and blockade spheres (red region). Ground state atoms (grey dots) in the blue region are subject to off-resonant laser coupling. (d):~Schematic phase diagram for the high (top) and low (bottom) temperature regimes depending on the total gas density $n$.}
\label{fig:intro}
\end{figure}
This model can be expanded to include a recovered state, from which an individual cannot be infected again, which corresponds to a susceptible-infected-recovered (SIR) model \cite{harko2014exact, bjornstad_2020, beckley2013modeling}. For Rydberg systems this can be approximately achieved by allowing an additional decay channel from the Rydberg state to a dark state $\ket{0}$ (see Fig.~\ref{fig:intro}a). A concrete mapping of the Rydberg facilitation system to SIS and SIR models will be discussed later.

The dynamics of both SIS and SIR models are well understood in the homogeneous mean-field regime.
While the homogeneous mixing assumption is well justified in systems where the infection spreading occurs on a regular lattice in high spatial dimensions, it fails in many relevant cases, for example if the the SIS/SIR dynamics take place on real-life networks. For such network systems, including e.g. random Erdős–Rényi (ER) \cite{erdHos1960evolution} and scale-free (SF) networks \cite{barabasi2009scale}, a large body of more sophisticated approximation methods have been developed \cite{pastor2001epidemic, pastor2002epidemic, duron2022mean}, but many questions remain unsolved. Here, Rydberg atoms provide a platform to experimentally investigate the epidemic dynamics on a variety of complex networks, which can e.g. be engineered by the use of tweezer arrays \cite{browaeys2020many}. Moreover, in a gas of atoms contained in some macroscopic trapping potential one can investigate the transition between a random ER network at very low temperatures, where the motion of atoms on the relevant time-scales of the facilitation process can be ignored, to the homogeneous mixing limit for a gas of high temperatures \cite{brady2023griffiths}. The latter corresponding to an annealed random network. 

With the addition of a slow decay channel into the inert (recovered) state, the loss of individuals in the population (corresponding to the sum of Rydberg and ground state densities) gradually moves the system into an absorbing-state. This gives rise to three typical epidemic stages, which have been experimentally observed in Rydberg facilitation systems \cite{natcom_griffiths}. Following an initial infection, there is a rapid epidemic growth in infected individuals, or Rydberg atoms, in the system. This is followed by a saturation and an eventual relaxation at long times as a result of the system reaching an absorbing-state on these time scales (see Fig.~\ref{fig:intro}b).

In order to describe the macroscopic dynamics of the Rydberg facilitation process in a gas, a simple mean-field model has been put forward in \cite{helmrich2020}, which however fails to provide a quantitative prediction of the microscopic dynamics, accurately calculated by Monte-Carlo simulations \cite{brady2023griffiths} (see Fig.~\ref{fig:modified-facilitation}). This discrepancy results from the mean-field model not regarding (i) Rydberg blockade, which prevents the excitation of any atom closer than some radius $r_\mathrm{blockade}$ to a Rydberg atom, and (ii) the emergent ER network at low temperatures.

In the following, we will develop a mean-field description of the dynamics of Rydberg excitations in a many-body facilitated gas that accounts for both of these effects and provides accurate predictions of the full facilitation dynamics, which we demonstrate by comparing our predictions with Monte-Carlo simulations. The network structure of the cold gas leads to a higher total gas density at very long times and subsequently a higher Rydberg density in the saturation phase in comparison to mean-field predictions. Rydberg Blockade causes a significant modification of the facilitation (infection) rate if the density of Rydberg excited atoms (infected individuals) reaches some threshold value. Similarly to the effect of regulatory measures on the dynamics of epidemics ("lockdown"),  it limits the maximum density of Rydberg-excited atoms (infected individuals), but at the same time leads to a substantial prolongation of the slow transition into the absorbing (recovery) phase.



\section{Microscopic Model of Rydberg facilitation}

A microscopic description of Rydberg facilitation in a gas can be achieved from a Lindblad master equation of the density matrix $\hat{\rho}$ which takes the form
\begin{align}
    \label{eq:master_equation}
    \ddt \hat{\rho} = i [\hat{\rho}, \hat{\mathcal{H}}] 
    + \sum_l \hat{L}_l \hat{\rho} \hat{L}_l^\dagger
    - \frac{1}{2} \{ \hat{L}_l^\dagger \hat{L}_l, \hat{\rho} \},
\end{align}
with the atom-light interaction Hamiltonian $\hat{\mathcal{H}}$ given by
\begin{align}
    \hat{\mathcal{H}} &= \sum_i
        \Big[
            \Omega (\sgr_i + \srg_i)
            +
            \Big(
                \sum_{j \neq i} \frac{c_6}{r_{ij}^6} \srr_j - \Delta
            \Big)
            \srr_i
        \Big].
\end{align}
Here $\hat{\sigma}_i^{\mu\nu}=\vert \mu\rangle_{ii}\langle \nu\vert$ is the projection operator of the $i$-th atom from the internal state $\nu$ to $\mu$. The external driving field is described by the Rabi-frequency $\Omega$ and the detuning $\Delta$, and the van-der-Waals interaction energy between Rydberg atoms $i$ and $j$ is given by ${c_6 / r_{ij}^6}$, with $r_{ij} $ being the distance between the atoms. Finally, in eq.~\eqref{eq:master_equation} dissipation is described by the jump operators $\hat{L}_l$. These take the form ${\hat{L}_1^{(i)} = \sqrt{(1 - b) \gamma}\hat{\sigma}_i^{gr}}$, ${\hat{L}_2^{(i)} = \sqrt{b \gamma}\hat{\sigma}_i^{0r}}$ for spontaneous decay from $\ket{r}$ to $\ket{g}$ and $\ket{0}$, as well as ${\hat{L}_3^{(i)} = \sqrt{\gamma_\perp}\hat{\sigma}_i^{rr}}$ for the dephasing of the Rydberg state. The parameter ${b \in [0, 1]}$ corresponds to the percentage of Rydberg atoms which spontaneously decay to $\ket{0}$ and are thereby removed from the system. Therefore, if ${b=0}$ the system corresponds to a two-level system and resembles an SIS epidemic (see Fig.~\ref{fig:intro}a). 

Dephasing results from e.g. Doppler broadening or the spread of the atomic wave packet over the van-der-Waals potential \cite{helmrich2020, motional_dephasing_lesanovsky}. In the large dephasing limit, the dynamics of a many-body Rydberg gas are effectively governed by classical rate equations \cite{Levi_2016}. As a result, this system can be simulated to great accuracy using Monte-Carlo simulations. Starting from eq.~\eqref{eq:master_equation}, after adiabatic elimination of coherences, one can formulate a set of rate equations for the probabilities of atom~$i$ being in the Rydberg state with $P_\mathrm{r}^{(i)}$ or ground state with $P_\mathrm{g}^{(i)}$ as
\begin{subequations}
    \label{eq:rate_equations}
    \begin{align}
        \label{eq:rate_equations_ryd}
        \ddt P_\mathrm{r}^{(i)} &= \gfac(\Sigma) P_\mathrm{g}^{(i)} - \bigl(\gfac(\Sigma) + \gamma\bigr) P_\mathrm{r}^{(i)}
        \\
        \label{eq:rate_equations_grd}
        \ddt P_\mathrm{g}^{(i)} &= \bigl(\gfac(\Sigma) + (1- b)\gamma\bigr)  P_\mathrm{r}^{(i)} - \gfac(\Sigma) P_\mathrm{g}^{(i)},
    \end{align}
\end{subequations}
with the stimulated excitation rate given by
\begin{align}
    \label{eq:gamma_fac}
    \gfac(\Sigma) = \frac{2 \Omega^2 \gamma_\perp}{\gamma_\perp^2 + \Delta^2
    \big(
    \sum_{\substack{j \neq i \\ j \in \Sigma}} \frac{\rfac^6}{r_{ij}^6} - 1
    \big)^2}.
\end{align}
Here $\Sigma$ is the set of indices of Rydberg-excited atoms. If no other Rydberg atom exists in the gas or their distance is much larger than $\rfac$, $\gfac(\Sigma)$ reduces to the off-resonant excitation rate of an isolated atom
\begin{align}
    \tau = \frac{2 \Omega^2 \gamma_\perp}{\gamma_\perp^2 + \Delta^2}.
\end{align}
If a Rydberg atom is present in the system, atoms located around a certain distance to it, called facilitation distance $\rfac$, are shifted into resonance and can be excited on a much faster time-scale, given by the facilitation rate ${\gfac = 2\Omega^2 / \gamma_\perp}$. The facilitation distance is given by
\begin{align}
    \rfac = \sqrt[6]{\frac{c_6}{\Delta}}.
\end{align}
Rydberg facilitation can be observed when ${\Delta \gg \Omega}$, as this naturally gives rise to a hierarchy in time-scales such that
\begin{align}
    \label{eq:hierarchy_timescales}
    \gfac \gg \gamma \gg \tau.
\end{align}
In this case, off-resonant excitations and the decay of Rydberg atoms are effectively static on the time-scale of facilitated excitations. Each Rydberg atom spans a spherical shell, with distance $\rfac$ and approximate width $\drfac$, in which atoms are resonantly coupled to the driving laser field.  The width of the facilitation shell is thereby determined by the effective linewidth of the excitation transition and reads
\begin{equation}
    \drfac = \frac{\gamma_\perp}{2 \Delta} \rfac.
\end{equation}
Atoms closer than $\rfac-\drfac/2$ to a Rydberg atom are subject to  Rydberg blockade \cite{lukin2001dipole} and cannot be excited, since they are shifted out of resonance again. These three regions around a Rydberg atom (off-resonant coupling, facilitation, and blockade) can be schematically seen in Fig.~\ref{fig:intro}c.

In \cite{helmrich2020}, a mean-field equation for a macroscopic description of the many-body Rydberg facilitation dynamics has been derived. For a homogeneous gas this reads
\begin{subequations}
    \begin{align}
        \label{eq:langevin_rho}
        \ddt \rho &= 
        -\kappa (2 \rho^2 - \rho n)
        -\gamma \rho
        -\tau (2\rho - n),
        \\
        \label{eq:langevin_n}
         \ddt  n &= - b\gamma  \rho.
    \end{align}
    \label{eq:langevin_diehl}
\end{subequations}
Here $\rho$ corresponds to the coarse grained Rydberg density (in a small volume $\Delta V$)
\begin{align}
    \label{eq:coarse_grain_rho}
    \rho(\vec{r},t) = \frac{1}{\Delta V}\sum_{i:\vec{r}_i\in \Delta V} \langle \srr_i\rangle,
\end{align}
and $n$ is the density of ground and Rydberg state atoms
\begin{align}
    \label{eq:coarse_grain_n}
    n(\vec{r},t) = \frac{1}{\Delta V}\sum_{i:\vec{r}_i\in \Delta V} \Bigl(\langle \srr_i\rangle + \langle \sgg_i\rangle\Bigr).
\end{align}
Note that $n$ does not count $\ket{0}$ state atoms and therefore decreases over time if ${b>0}$. The spreading rate of Rydberg excitations in the many-body gas is given by the two-body facilitation rate integrated over the facilitation shell
\begin{align}
    \kappa = \gfac V_s,    
\end{align}
with the volume of the facilitation shell ${V_s \approx 4 \pi \drfac \rfac^2}$. The above equations predict an absorbing-state phase transition between an active and absorbing phase for the critical gas density
\begin{align}
    \label{eq:n_crit}
    n_\mathrm{crit} = \frac{\gamma}{\gfac V_s}.
\end{align}
Equation~\eqref{eq:langevin_rho} for the (mean-field) Rydberg density in the many-body gas strongly resembles the SIS equation of motion of infected individuals given by \eqref{eq:rhoI}. However, in contrast to SIS/SIR epidemics, Rydberg systems additionally feature (i) off-resonant excitations with rate $\tau$, (ii) resonant (facilitated) \textit{de}-excitations of Rydberg atoms (described by the term ${-2\kappa \rho^2}$ in eq.~\eqref{eq:langevin_rho}), and (iii) Rydberg blockade which is not regarded in eq.~\eqref{eq:langevin_rho}.
 Thus eqs.~\eqref{eq:langevin_diehl} do not directly map to those of the SIS dynamics eqs.(1) (with ${b=0}$). However, identifying $\rho^\textrm{I}= 2\rho$ and $\rho^\textrm{S}= n- 2\rho\ge 0$ and setting $\tau=b=0$ reproduces the SIS equations with conserved total density $\rho^\text{I}+\rho^\text{S}=n-2\rho +2 \rho = n$.  Moreover in the high temperature limit, the de-excitation can effectively be neglected as it is a second order process in terms of Rydberg density. 

For the low temperature gas, the excitations dynamics are constrained to an ER network in which the individual nodes are comprised of atoms (either in $\ket{g}$ or $\ket{r}$), and connections between nodes, say $i$ and $j$ exist if ${r_{ij} \approx \rfac}$. The number of connections a node has (i.e. the number of atoms with distance $\rfac$ to an atom) is called the degree $k$ of the atom. In an ER network the node degrees follow a Poissonian distribution with average degree
\begin{align}
    \label{eq:avg_degree}
    \langle k \rangle = n V_s.
\end{align}
This system features a percolation transition at ${\langle k \rangle = 1}$ between an (almost fully) connected network and a fragmented network, comprised of many, small disconnected clusters. Here, clusters refers to a group of connected nodes. From eq.~\eqref{eq:avg_degree}, we can identify the gas density at which the percolation transition occurs as 
\begin{align}
    \label{eq:n_perc}
    n_\mathrm{perc} = \frac{1}{V_s}.
\end{align}
This density is a factor ${\gfac / \gamma}$ larger than the critical density of the phase transition to the absorbing phase for a homogeneous gas \cite{brady2023griffiths}. A schematic phase diagram for the high and low temperature gas can be seen in Fig.~\ref{fig:intro}d.

In this paper, we model the actual many-body dynamics using Monte-Carlo simulations of the rate equations~\eqref{eq:rate_equations}. We assume a cubic box with length ${L = 7 \rfac}$ and periodic boundary conditions. Atom positions are chosen randomly and velocities are sampled from a Maxwell-Boltzmann distribution with temperature parameter $\hat{v}$, corresponding to the most probable atom velocity in the gas. For the time evolution we utilize a fixed time step Monte-Carlo algorithm \cite{RUIZBARLETT20095740}, with the time step ${dt = 1/400 \, \gamma^{-1}}$, as long-range interactions paired with the fast movement of atoms in the high temperature case results in quickly changing transitional rates in the system. To ensure numeric stability the $c_6$ potential in eq.~\eqref{eq:gamma_fac} is truncated at a cutoff value around the singularity ${r_{ij} \to 0}$.

\section{Modified Langevin Description of Epidemic Evolution}

In the following section, we develop an effective macroscopic theory of the Rydberg facilitation process, expanding the Langevin equation~\eqref{eq:langevin_diehl}, starting from the microscopic model. This new equation takes into account Rydberg blockade, as well as the network structure in the case of the low temperature gas.

\begin{figure}[H]
    \centering
    \includegraphics[width=\columnwidth]{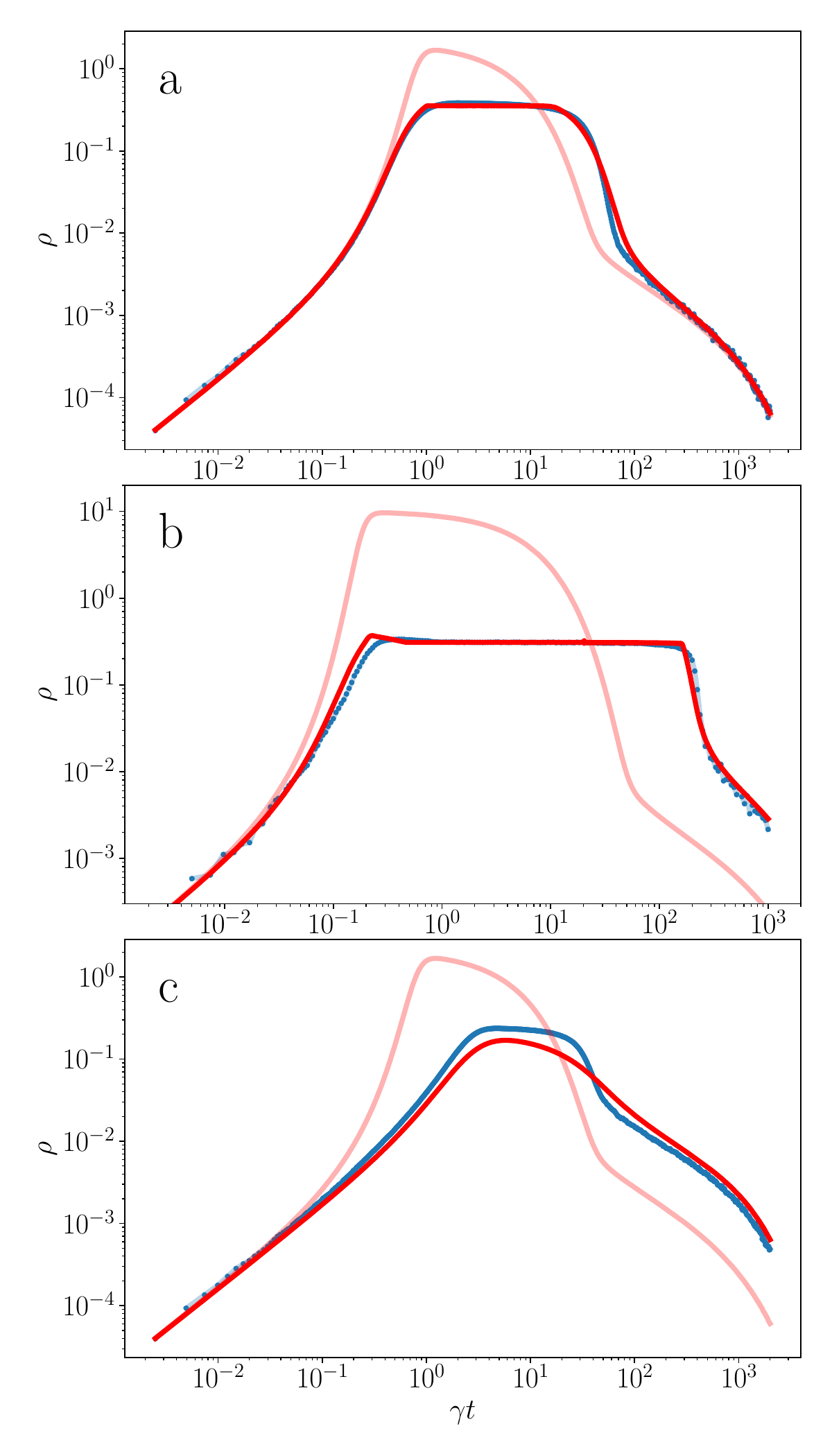}
    \caption{Rydberg density $\rho$ over time for ${\rho(t=0) = 0}$ and ${b=0.3}$ modeled with Monte-Carlo simulations (blue dots), eq.~\eqref{eq:langevin_diehl} (red, faint), and eq.~\eqref{eq:langevin_rho_full} (red, solid), for
    \textbf{a} high temperature gas with starting density ${n_0 = 4.0 \, \rfac^{-3} > n_\mathrm{crit}}$, \textbf{b} low temperature percolating gas  with ${n_0 = 20 \, \rfac^{-3} > n_\mathrm{perc}}$, and \textbf{c} low temperature non-percolating gas  with ${n_0 = 4 \, \rfac^{-3} < n_\mathrm{perc}}$.}
    \label{fig:modified-facilitation}
\end{figure}

In Fig.~\ref{fig:modified-facilitation}, the dynamics of Rydberg excitations predicted by the improved Langevin equation and by Monte-Carlo simulations are compared for the cases: (a) high temperature gas with ${n_0 > n_\mathrm{crit}}$, (b) low temperature gas initially above the percolation threshold with ${n_0 > n_\mathrm{perc}}$, and (c) low temperature gas initially below the percolation threshold ${n_0 < n_\mathrm{perc}}$. Here $n_0$ refers to the gas density at ${t=0}$. Additionally, we use a branching parameter ${b=0.3}$, allowing some loss into the recovered state~$\ket{0}$. Therefore, for all cases (a-c), the system drives itself to the absorbing-state and follows the typical epidemic stages as seen in Fig.~\ref{fig:intro}b.

We start from the microscopic Heisenberg-Langevin equations describing the quantum many-body dynamics of Rydberg excitations for atoms at given spatial positions given by
\begin{align}
    \ddt \srr_i &= 
    -i\Omega (\srg_i - \sgr_i) - \gamma \srr_i + \hat{\xi}_1,
    \\
    \ddt \srg_i &=
    -i
    \Big(
        \Omega (\srr_i - \sgg_i) - \hat{V}_i \srg_i
    \Big)
    - \gamma_\perp\srg_i + \hat{\xi}_2.
\end{align}
We have accounted for losses and dephasing and added corresponding fluctuation operators $\hat{\xi}_1$ and $\hat{\xi}_2$, which disappear in the quantum mechanical average and whose properties can be obtained from  the fluctuation-dissipation relation \cite{kubo1966fluctuation}. 
The operator $\hat{V}_i$ describes the detuning of the $i$-th atom and depends on the states of all other atoms. It is given by
\begin{align}
    \label{eq:potential_V_k}
    \hat{V}_i = \Delta 
    \Big( 
        -1 + \sum_{j \neq i} \frac{\rfac^6}{r_{ij}^6}
        \srr_j
    \Big).
\end{align}
We note that the operator valued quantities are objects in Hilbert space
describing the quantum mechanical evolution and are subject to the classical statistics of the (time dependent) random positions of the atoms.
The dynamics of the atom positions are treated classically, which is well justified in the high-dephasing limit, assumed 
throughout the present paper. 

Furthermore, the effect of $c_6$ forces acting on the center-of-mass 
motion of the atoms due to the distance dependence of $\hat{V}_i$ are disregarded in the present paper. They will be discussed elsewhere in more detail \cite{Daniel-2023b}, where we will show that under typical experimental conditions they can be accounted for by a change of the atoms velocity distribution and, in the case of a trapped gas, by an additional loss channel.

Assuming high dephasing ${\gamma_\perp \geq \Omega}$, the coherences $\srg_i$ quickly decay to quasi-stationary values relative to the relevant many-body time scales. Therefore, we adiabatically eliminate coherences (${\ddt \srg_i = 0}$) and arrive at
\begin{align}
    \label{eq:dsrrdt-full}
    \ddt \srr_i &=-
    \frac{2\Omega^2 {\gamma_\perp}}{{\gamma_\perp}^2 + \hat{V}_i^2}(\srr_i - \sgg_i) - \gamma \srr_i + \hat{\xi}.
\end{align}
In the following, we expand the leading fraction in eq.~\eqref{eq:dsrrdt-full} with a full basis of projection operators ${\Pi}_i(m)$ projecting onto $m$ Rydberg atoms in the facilitation \textit{sphere} of atom $i$. Each of these atoms have a relative distance to atom $i$ in the interval $0 \leq {r_{ij} \leq \rfac, \; \forall j \in [1, m]}$. Using this, the equation of motion takes the form
\begin{align}
    \nonumber
    \ddt \srr_i &= 
    -{\Pi}_i(0) \frac{2\Omega^2{\gamma_\perp}}{{\gamma_\perp}^2 + \Delta^2} (\srr_i - \sgg_i)
    \\
    \nonumber
    &- {\Pi}_i(1) 
    \underbrace{
    \frac{2\Omega^2 {\gamma_\perp}}{{\gamma_\perp}^2 + \Delta^2
    \big(
        (
            \frac{\rfac}{r_{1i}}
        )^6 - 1
    \big)^2
    }
    }_{(*)}
    (\srr_1 \srr_i - \srr_1 \sgg_i)
    \\
    \nonumber
    &+ ...
    \\
    &- \gamma \srr_i + \hat{\xi}.
    \label{eq:ddt_srr_expanded}
\end{align}
All rates for more than one Rydberg atom in the facilitation sphere ${m > 1}$ are strongly suppressed due to blockade. As a result, we truncate the expansion at ${m=1}$.

The fraction $(*)$ in eq.~\eqref{eq:ddt_srr_expanded} is the distance dependent facilitation rate for an atom in the presence of one Rydberg atom. Here, we approximate this rate as being non-zero only if atom $i$ is in the facilitation \emph{shell} of the Rydberg atom ${r_{1i} \in S_\text{f} \equiv [\rfac - \frac{\delta \rfac}{2}, \rfac + \frac{\delta \rfac}{2}]}$. We express this as

\begin{align}
  \srr_1  \Pi_i(1) \to \srr_1 \, \Xi_i(1) = \srr_1 \times
    \begin{cases}
        1, \quad  r_{1i} \in S_\text{f} \\
        0, \quad \text{else}
    \end{cases},
\end{align}
and obtain
\begin{align}
    \nonumber
    \ddt \srr_i &= 
    -\tau\, {\Pi}_i(0)  (\srr_i - \sgg_i)
    \\
    \nonumber
    &-  \gfac \, {\Xi}_i(1)
    (\srr_1 \srr_i - \srr_1 \sgg_i)
    \\
    &- \gamma \srr_i + \hat{\xi},
\end{align}
with the maximal facilitation rate $\gfac = \frac{2 \Omega^2}{{\gamma_\perp}}$ and the off-resonant excitation rate $\tau = \frac{2\Omega^2 \gamma_\perp}{{\gamma_\perp}^2 + \Delta^2}$. 

Finally, we calculate the expectation value of the operator $\srr_i$ with a double averaging over the quantum mechanical state and the ensemble  of the many different atom positions in the gas. We will denote these
double averages as $\langle \langle \srr_i \rangle \rangle$ and write
\begin{align}
    \nonumber
    \ddt \langle \langle \srr_i \rangle \rangle &= 
    -\tau \langle \langle{\Pi}_i(0) (\srr_i - \sgg_i) \rangle \rangle
    \\
    \nonumber
    &- \gfac \, 
    \langle \langle{\Xi}_i(1) \srr_1
    (\srr_i - \sgg_i) \rangle \rangle
    \\
    &- \gamma
    \langle \langle \srr_i \rangle \rangle
    \\
    \nonumber
    &\approx
    -\tau \langle \langle {\Pi}_i(0) \rangle \rangle
    \Bigl(\langle \langle \srr_i \rangle \rangle - \langle \langle \sgg_i \rangle \rangle\Bigr)
    \\
    \nonumber
    &- \gfac \, 
    \langle \langle {\Pi}_i(1) \srr_1 \rangle \rangle
    \Bigl(\langle \langle \srr_i \rangle \rangle - \langle \langle \sgg_i \rangle \rangle\Bigr)
    \\
    &- \gamma
    \langle \langle \srr_i \rangle \rangle.
\end{align}
Assuming a randomly distributed gas, we can approximate the probabilities $\langle \langle {\Pi}_i(m) \rangle \rangle$ as Poissonian with the rate $\rho V_\textrm{f}$, i.e. $\langle \langle {\Pi}_i(m) \rangle \rangle = (\rho V_f)^m e^{-\rho V_f}/m!$, 
resulting in
\begin{align}
    \langle \langle {\Pi}_i(0) \rangle \rangle 
    &= 
    \text{e}^{-\rho V_\textrm{f}} 
    \\
    \langle \langle {\Xi}_i(1) \srr_1 \rangle \rangle 
    &\equiv \frac{V_s}{V_f}
    \langle \langle {\Pi}_i(1) \rangle \rangle
    = 
    \rho V_\textrm{s} \, \text{e}^{-\rho V_\textrm{f}}.
\end{align}
The factor $V_s/V_f$ in the second line takes into account that only atoms in the facilitation shell contribute.  We then perform the coarse-graining given by eqs.~\eqref{eq:coarse_grain_rho}~and~\eqref{eq:coarse_grain_n} and arrive at
\begin{align}
    \label{eq:langevin_rho_almost_full_high_temp}
        \nonumber
        \ddt \rho = 
        &-\kappa \text{e}^{-\rho V_\text{f}} \rho (2 \rho - n)
        \\
        &-\gamma \rho
        -\tau (2\rho - n).
\end{align}
Furthermore, we have assumed here ${\text{e}^{-\rho V_\text{f}} \tau \approx \tau}$ as the off-resonant rate is only relevant when ${\rho V_\text{f} \ll 1}$.

The spreading rate of excitations $\kappa$ is now exponentially damped by the density of Rydberg atoms. This gives a better description of the spreading of Rydberg excitations in the epidemic growth stage. However, as all atoms with distances closer than $r_\mathrm{blockade}$ to a Rydberg atom cannot be excited due to Rydberg blockade (red region in Fig.~\ref{fig:intro}c), there exists a maximum density of Rydberg atoms $\rho_\mathrm{max}$, given by the packing density of non-overlapping spheres, above which no more excitations are possible.

To quantify the blockade induced saturation density in the gas, we introduce the parameter $\eta$ corresponding to the packing density of spheres in a given volume.

For the high temperature gas, this corresponds to the densest packing of spheres, given by ${\eta = \frac{\pi}{3\sqrt{2}} \approx 74.0 \%}$. In this regime we can assume this packing density to be achieved, as the high motion of the atoms allow the system to organize itself to this state.

For the frozen gas, the packing density is slightly lower, and is given by the closest density of \textit{randomly} packed spheres, which is given by ${\eta \approx 63.5 \%}$ \cite{sphere_packing}.

As $\drfac \ll \rfac$ we can approximate the blockade radius as ${r_\mathrm{blockade} \approx \rfac}$ and write
\begin{align}
    \label{eq:rho_max}
    \rho_\mathrm{max} = 2 \frac{\eta}{V_\mathrm{fac}},
\end{align}
with the approximate volume of the blockade sphere ${V_\mathrm{fac} = \frac{4}{3} \pi \rfac^3}$. The factor of 2 emerges as when a facilitation event occurs, the facilitated atom is centered on the blockade sphere of the facilitating Rydberg atom. As a result, the blockade spheres of these atoms overlap. If, however, a third Rydberg atom is facilitated (by the second Rydberg atom), its blockade sphere borders the blockade sphere of the first Rydberg atom with, on average, very little overlap (see Fig.~\ref{fig:intro}c).

As the laser coupling smoothly changes from resonant, for an atom with distance ${r = \rfac}$ to a Rydberg atom, to strongly suppressed for ${r < \rfac}$, this can be regarded as a packing of soft spheres with an uncertainty in volume of ${\delta V_\mathrm{fac} = 4\pi \drfac \rfac^2}$. The result is a smearing out of $\rho_\mathrm{max}$ given by ${\delta \rho_\mathrm{max} = \delta V_\mathrm{fac} \frac{2 \eta}{V_\mathrm{fac}^2}}$. We can now add a heuristic function which sets the facilitation rate to 0 if ${\rho > \rho_\mathrm{max}}$ as
\begin{align}
    h(\rho) = \frac{1}{2} 
    \Big(
        1 + \tanh
        \Big(
            \frac{\rho_\mathrm{max} - \rho}{\delta \rho_\mathrm{max}}
        \Big)
    \Big)
    .
\end{align}

The added factors $\text{e}^{-\rho V_f}$ and $h(\rho)$ to the facilitation rate~$\kappa$ suffice to accurately describe the dynamics of the Rydberg density in the high temperature gas (see Fig.~\ref{fig:modified-facilitation}a).

For the low temperature gas the finite connectivity greatly reduces the facilitation rate. Taking into account that facilitation can only occur if the degree of the atom $k$ is not $0$, we alter the facilitation rate to
\begin{align}
    \label{eq:fractured_gamma_fac}
    \kappa \to \kappa \, P(k > 0).
\end{align}
For an ER network with average degree $\langle k \rangle \ll 1$, we can approximate $P(k > 0) \approx \langle k \rangle$. In this case the new infection rate $\kappa$ corresponds to the Kephart-White model \cite{kephart1992directed, chakrabarti2008epidemic}.

The full Langevin equation for the Rydberg density  reads
\begin{align}
    \label{eq:langevin_rho_full}
        \nonumber
        \ddt \rho = 
        &-\kappa 
        \text{e}^{-\rho V_\text{f}} 
        h(\rho)
        P(k > 0) \rho (2 \rho - n)
        \\
        &-\gamma \rho
        -\tau (2\rho - n),
\end{align}
with ${P(k>0) = 1 - \text{e}^{-n V_s}}$ for the low temperature gas and ${P(k>0) = 1}$ at high temperatures.

In Fig.~\ref{fig:modified-facilitation} we have compared the predictions from the modified Langevin equation~\eqref{eq:langevin_rho_full} with Monte Carlo simulations in the high temperature gas, the frozen percolating gas, and the frozen non-percolating gas. For the high temperature and the frozen percolating case, eq.~\eqref{eq:langevin_rho_full} has a very good agreement for all epidemic stages with Monte-Carlo data. In particular, it predicts the correct density in the saturation stage in the high temperature and the low temperature percolating gas in contrast to eq.~\eqref{eq:langevin_diehl}.

Furthermore, for the case of the low temperature gas, eq.~\eqref{eq:langevin_rho_full} gives a much better approximation of the relaxation epidemic stage (i.e. for times ${\gamma t \gtrsim 10^2}$). In this stage, the Rydberg density is much higher than the expected MF density (predicted by the faint red line), which holds for high temperatures. In contrast to the high temperature regime, the system leaves the active phase at much higher gas density due to the finite connectivity of excitation paths in the gas. The factor~${P(k>0)}$ in the facilitation rate gives a much better approximation of this increased Rydberg density.

\section{Conclusion}

In conclusion, we have developed a modified mean-field approach to model the Rydberg density over time in a many-body gas under  facilitation conditions for the limits of high and low temperature. In the low temperature regime, we have additionally differentiated between a system with initial density ${n_0 > n_\mathrm{prec}}$ and ${n_0 < n_\mathrm{perc}}$,  where $n_\mathrm{perc}$ is the percolation density below which heterogeneous effects play a large role.

Our modelling is similar to that developed in \cite{helmrich2020}, but with three key improvements to the facilitation (or infection) rate $\kappa$. We consider (i) random atom positions leading to a Poissonian distribution in the number of Rydberg atoms closer than $r_\mathrm{blockade}$ to a given atom. In this case, the atom cannot be excited or de-excited due to Rydberg blockade. As a result, with increasing Rydberg density, the global facilitation rate $\kappa$ exponentially decreases.

Additionally, (ii) excited Rydberg atoms can be seen as soft spheres inside of which no atoms can be excited due to blockade. Therefore, there exists a tightest packing of excited atoms beyond which the facilitation rate~$\kappa$ vanishes. In the high temperature regime, this packing density corresponds to the tightest packing of spheres in a given volume, as the high thermal velocities allow the system to continuously organize itself to this state. In the low temperature regime the packing density is given by that of randomly placed spheres in a given volume, which, in comparison, is slightly lower. 

Finally, (iii) for the low temperature regime, one has to additionally consider the finite connectivity of the underlying network along which facilitated excitations can spread. On a mean-field level, we have described this by reducing the facilitation rate in correspondence with the portion of atoms with network degree (i.e. the number of atoms in their facilitation shell) ${k = 0}$. The percentage of these isolated atoms increases as the network connectivity decreases, and is therefore dependent on the total density of the gas.

For both the high temperature, as well as the low temperature, high density case eq.~\eqref{eq:langevin_rho_full} gives excellent correspondence to Monte-Carlo data for all epidemic stages.

For the low temperature, low density gas the system is characterized by strong heterogeneity making an accurate mean-field description challenging. However, for this case we still see a large improvement in the Langevin description of the dynamics.

\subsection*{Acknowledgement}

The authors thank Simon Ohler and Johannes Otterbach for fruitful discussions. 
Financial support from the DFG through SFB TR 185, project number
277625399, is gratefully acknowledged.

\bibliography{references}

\end{document}